\documentclass [aps,prl,superscriptaddress,twocolumn,showpacs,a4paper,unsortedaddress,
amsmath,amssymb,byrevtex,floatfix]{revtex4}

\usepackage[latin1]{inputenc}
\usepackage{graphicx}
\usepackage{dcolumn}
\usepackage{bm}
\usepackage{hyperref}
\usepackage{times}

\begin{document}

\title{Non-equilibrium transport response from equilibrium transport theory}

\author{V. M. Garc\'{\i}a-Su\'arez}
\email{vm.garcia@cinn.es}
\affiliation{Departamento de F\'{\i}sica, Universidad de Oviedo, 33007 Oviedo, Spain}
\affiliation{Nanomaterials and Nanotechnology Research Center (CINN, CSIC - Universidad de Oviedo), Spain}
\affiliation{Department of Physics, Lancaster University, Lancaster LA1 4YB, United Kingdom}

\author{J. Ferrer}
\affiliation{Departamento de F\'{\i}sica, Universidad de Oviedo, 33007 Oviedo, Spain}
\affiliation{Nanomaterials and Nanotechnology Research Center (CINN, CSIC - Universidad de Oviedo), Spain}
\affiliation{Department of Physics, Lancaster University, Lancaster LA1 4YB, United Kingdom}

\date{\today}

\begin{abstract}
We propose a simple scheme that describes accurately essential non-equilibrium effects in nanoscale electronics devices using equilibrium
transport theory. The scheme, which is based on the alignment and dealignment of the junction molecular orbitals with the shifted Fermi
levels of the electrodes, simplifies drastically the calculation of current-voltage characteristics compared to typical non-equilibrium
algorithms. We probe that the scheme captures a number of non-trivial transport phenomena such as the negative differential resistance
and rectification effects. It applies to those atomic-scale junctions whose relevant states for transport are spatially placed on
the contact atoms or near the electrodes.
\end{abstract}


\maketitle
Nano-scale devices tend to suffer from a serious lack of reproducibility from one research group to another, and from a lack of durability
and robustness\cite{ITRS10,Heath09}. However, the consistent and thorough work of a range of experimental groups have enabled to
reach some consensus on the conductance values and variability of a few specific junctions\cite{Bergren09,Poulson09,Cuevasbook}. In addition,
the development of simulation codes based on a combination of density functional theory (DFT) and the non-equilibrium Green's function
formalism (NEGF)\cite{Cuevasbook,Dattabook,NEGF1,NEGF2,NEGF3,NEGF4,NEGF5,NEGF6} have enabled to assist the above experiments with theoretical
insights and predictions. However, the size and complexity of the experimentally active part of the junction is typically too large,
and the above codes need to assume a number of simplification regarding the size, geometry, number of feasible atomic arrangements
and the electronic correlations. Recently, some of the above codes have been combined with classical force field programs, enabling
the simulation of much larger systems and the sampling of a much wider set of atomic
arrangements\cite{Pablo11,Colin11}. Nevertheless, present current-voltage $I-V$ characteristics are still computed using NEGF
techniques, which are rather cumbersome and extremely computer hungry. In other words, the NEGF is a serious bottleneck hindering
the deployment of realistic transport simulations at the nanoscale. In contrast, equilibrium transport techniques are much simpler, better founded on
physical grounds and computationally drastically less demanding\cite{Dattabook}. However, they are completely unable to reproduce current-voltage
curves of nano-scale devices\cite{Comparison}. This is so because these techniques assume that the electronic states and the Hamiltonian at
the junction do not change under the application of a voltage. However, a voltage bias drives a flow of electrons into and out of the junction,
driving the device out of equilibrium and possibly even changing its quantum state and sometimes even its atomic arrangement. For example,
the electric field originated by the voltage bias can also shift the energy position of the molecular levels by the Stark effect, depending
on the polarizability of each molecular level or on whether the molecule has itself a polar nature\cite{Die07,Baa09}.

We show here a suitable modification of equilibrium transport techniques that accounts for a large set of non-equilibrium
transport phenomena such as negative differential resistance (NDR)\cite{NDR96} or rectification effects. The performance of our method
depends on the energy position, electrode coupling and spatial localization of the molecular orbitals that are responsible
for the transport properties. This works specially well for systems where the closest states to the Fermi level (HOMO or LUMO) are located at the contact atoms or near the electrodes. We discuss below the physical
mechanisms at work for several molecular-scale functions. Our scheme enables the  simulation of $I-V$ curves which compare
favorably to those obtained using NEGF techniques. As a consequence, we also propose that a number of transport phenomena
regarded up to now as genuine manifestations of non-equilibrium behavior of a junction could be recognized as quantitative
but not qualitative modifications of its equilibrium state.

We use Caroli's scheme \cite{Car72} and split the device into left and right electrodes, and a extended molecule. The extended molecule includes several principal layers of the electrodes, as well
as the molecule itself. The full Hamiltonian ${\cal H}$ can be written as

\begin{equation}
\left(\begin{array}{ccc}
{\cal H}_\mathrm{L}+\frac{eV}{2}{\cal S}_\mathrm{L}&{\cal H}_\mathrm{LM}+\frac{eV}{2}{\cal S}_\mathrm{LM}&0\\
{\cal H}_\mathrm{ML}+\frac{eV}{2}{\cal S}_\mathrm{ML}&{\cal H}_\mathrm{M}[\rho]&{\cal H}_\mathrm{MR}-\frac{eV}{2}{\cal S}_\mathrm{MR}\\
0&{\cal H}_\mathrm{RM}-\frac{eV}{2}{\cal S}_\mathrm{RM}&{\cal H}_\mathrm{R}-\frac{eV}{2}{\cal S}_\mathrm{R}\\
\end{array}\right)\:,
\end{equation}

\noindent where the matrices ${\cal H}_\mathrm{L}$, ${\cal
H}_\mathrm{R}$, ${\cal H}_\mathrm{LM}$, ${\cal H}_\mathrm{RM}$ and
their corresponding overlap matrix blocks, indicate respectively
the left- and right-hand side leads Hamiltonians and the coupling
matrices between the leads and the extended molecule.
The matrix elements of the extended molecule Hamiltonian ${\cal H}_\mathrm{M}^{i,j}$ run over all the
orbitals residing in the extended molecule region. They depend on the non-equilibrium density matrix
$\rho$, which must be calculated self-consistently for
each voltage $V$ using cumbersome and computationally very demanding NEGF techniques\cite{Cuevasbook,NEGF1,NEGF2,NEGF3,NEGF4,NEGF5,NEGF6}.

\begin{figure}
\includegraphics[width=0.9\columnwidth]{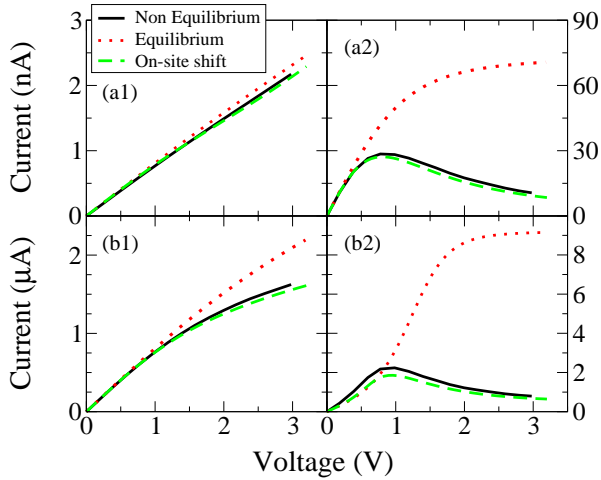}
\caption{\label{Fig1}(Color online) $I-V$ curves of a junction made of two carbon chains, each terminated by a nitrogen (a) or a sulfur atom (b).
These two atoms are separated by a physical gap of 3.7 \AA. Panels (a1) and (b1) correspond to fully relaxed coordinates. Panels (a2) and (b2) correspond to
chains where the terminating nitrogen and sulfur atoms have been separated by 0.5 \AA~ away from their relaxed positions. Solid and dotted 
lines correspond to NEGF and equilibrium simulations, respectively. Dashed lines correspond to simulations using the scheme proposed in this article. }
\end{figure}

We propose here that in many situations $\rho$ and its derived ${\cal H}_\mathrm{M}[\rho]$ need only be computed at
zero voltage, and that the effect of a finite bias can be accounted for by the alignment or dealignment of the
molecular orbitals at the extended molecule region with the energy levels of the electrodes (which are shifted by $\pm e\,V/2$ by the bias voltage).
Mathematically, we apply a simple shift to the Hamiltonian matrix elements from ${\cal H}_\mathrm{M}$ to ${\cal H}_\mathrm{M}+eV_i\,S_\mathrm{M}$,
where the local shifts $V_i$ depend on the junction nature. For example, in a highly-transparent junction, the shifts
can be modeled by a linear voltage ramp connecting the matrix elements of the orbitals at the outest layers. In contrast, most
nanoscale junctions contain a molecule that is connected to the left and right electrodes by weaker links, which means the details of the coupling of the
the electrodes to the molecular orbitals responsible for the transport must be taken into account.
Our scheme is specially suited for junctions where the HOMO and LUMO orbitals are spatially localized at the linker atoms joining
the molecule with the electrodes. In this case, we take $V_i=\pm e\,V/2$ for all orbitals starting at the outest layers in the extended molecule
and up to the linker atoms,  and $V_i=0$ for all the remaining inner orbitals. We show now how these shifts allow to mimic accurately a number
of simple and not so simple junctions displaying non-trivial negative differential resistance, as well as rectification effects for some
asymmetric molecules.

Notice that in NEGF simulations the left and right electrodes are defined as semi-infinite electron reservoirs so that a steady electron flow is established.
The chemical potentials at the left and right electrodes are shifted by $\pm eV/2$ and a permanent potential drop is produced somewhere in the extended
molecule. These features are reproduced by our scheme. We have also checked that they are not reproduced by applying an electric
field along the transport direction in equilibrium simulations of finite-sized systems that represent the extended molecule.
This is so because in these later cases the junction has a finite length and equilibrium conditions are reestablished at the end
of the simulation. We have verified that the applied electric field shifts initially the electronic states as in our scheme, but
because the whole device has a common Fermi energy, a positive charge surplus accumulates at the negative electrode and vice versa, thereby
screening the potential created by the electric field and moving back the energy levels towards their original positions.

\begin{figure}
\includegraphics[width=0.9\columnwidth]{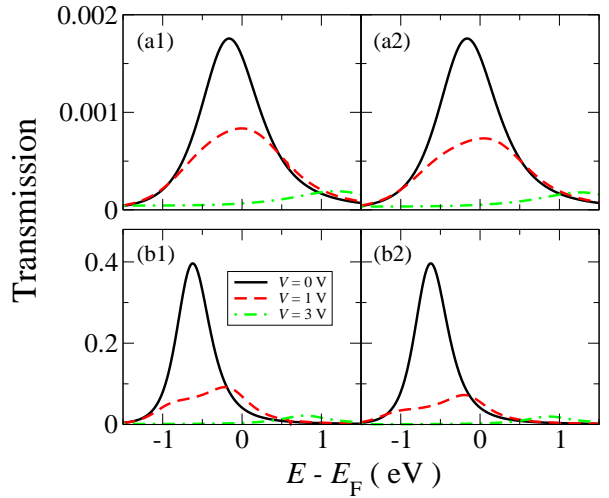}
\caption{\label{Fig2}(Color online) Transmission curves at different voltages for the carbon junctions shown in panels (a2) and (b2) of Fig. (\ref{Fig1}).
Panels (a1) and (b1) correspond to a full NEGF calculation. Panels (a2) and (b2) correspond to the scheme proposed in this article.}
\end{figure}

We discuss first two simple examples to illustrate the scheme. These systems are made of two semi-infinite carbon chains terminated each by either
a sulfur or a nitrogen atom and separated initially by a vacuum region of $3.7$ \AA. We apply a shift of $\pm e\,V/2$ to each Hamiltonian matrix
element at the extended molecule region up to and including the terminating atoms as proposed above. The left panels in Fig. (\ref{Fig1}) show
the $I-V$ characteristics of the two junctions, which display a conventional ohmic behavior. However, these junctions show NDR when
the terminating atoms are slightly displaced away from the chains\cite{Carrascal12}.
The $I-V$ curves of the two junctions are plotted in the right panels of Fig. (\ref{Fig1}). The NDR response shown is originated by a non-trivial energy shift
and splitting of the resonances appearing in the transmission coefficients $T(E)$ as the voltage bias is ramped up. This non-trivial evolution
of the transmission resonances is shown in the left panels of Fig. (\ref{Fig2}) and could be thought of a genuine non-equilibrium effect at first sight.
That this is so could also be inferred from the $I-V$ curves computed from the equilibrium equation

\begin{equation}\label{Eq2}
 I=\frac{e}{h}\int_{-e\,V/2}^{e\,V/2}\mathrm{d}E\,T(E)
\end{equation}

\noindent where $T(E)$ is calculated at zero voltage. Such $I-V$ curves are indeed quite different from those computed using standard NEGF techniques,
as shown in the right panels of Fig. (\ref{Fig1}). However, the figure also shows that Eq. (\ref{Eq2}) can produce
extremely accurate results if $T(E)$ is calculated using a modified Hamiltonian ${\cal H}_\mathrm{M}$ that follows our scheme.
Furthermore, the right panels in Fig. (\ref{Fig2}) show that the non-trivial bias evolution of the transmission resonances is also followed by
our scheme with high fidelity. As a consequence, we propose that the resonance splitting and shift shown in the figure is only apparent. What
actually happens is that the single resonance is cut-off by an opposite shift of the on-site energy levels at opposite sides of the junctions,
and is therefore not a genuine non-equilibrium effect.

The scheme proposed cannot be applied to all physical situations. An example where the scheme fails occurs when the molecular orbital responsible for transport
(HOMO or LUMO) is well localized in the central region of the molecule. In this case,
the resonance associated with that molecular orbital could be pinned to the Fermi level of either the left or the right electrode\cite{Gar08}. However, even in this scenario the scheme proposed improves slightly the $I-V$ curves compared to raw equilibrium simulations, because
it takes into account a non-equilibrium mechanism that is present in all nano-scale junctions.
We discuss next a series molecular junctions where different polyyne derivative are contacted by (001) gold electrodes to illustrate this argument.
Results for the synthesis of the molecules and the transport properties of the devices have been reported in Ref. [\onlinecite{Wang09}].

\begin{figure}
\includegraphics[width=0.9\columnwidth]{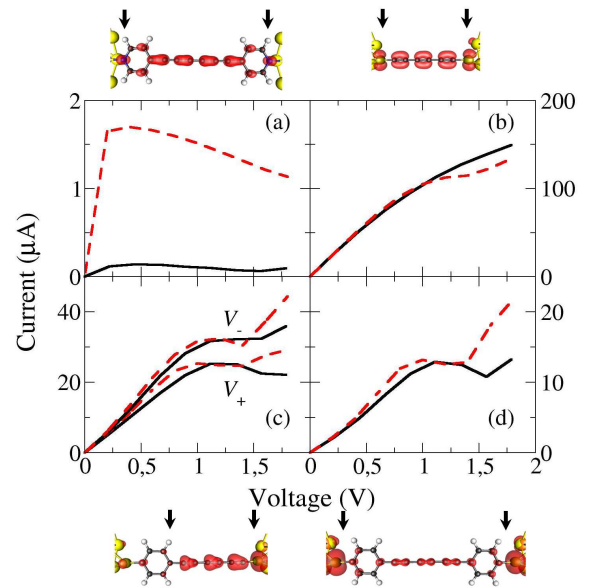}
\caption{\label{Fig3}(Color online) $I-V$ curves of gold (001) junctions sandwiched by pyridine-polyyne-pyridine (a), sulfur-polyyne-sulfur (b), sulfur-benzene-polyyne-sulfur (c) and sulfur-benzene-polyyne-benzene-sulfur (d) molecular wires,
where the polyyne has $n=3$ carbon pairs. The molecules are sketched above or below each corresponding graphic, together with density contours that
highlight the localization of the relevant molecular orbital. Solid lines correspond to non-equilibrium calculations, while dashed lines correspond
to calculations performed with the scheme proposed in this article. The on-site energy levels have been shifted up to the atoms indicated by the arrows.}
\end{figure}

\begin{figure}
\includegraphics[width=0.9\columnwidth]{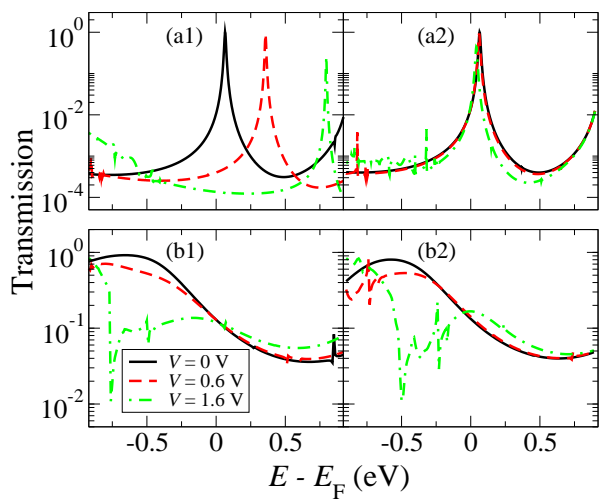}
\caption{\label{Fig4}(Color online). Transmission curves at different voltages for the pyridine-polyyne-pyridine (top panels) and
sulfur-benzene-polyyne-benzene-sulfur (bottom panels) junctions shown in Panels (a) and (c) of Fig. (\ref{Fig3}). Panels (a1) and (b1) correspond
to a full non-equilibrium calculation. Panels (a2) and (b2) correspond to the scheme proposed in this article.}
\end{figure}

Fig. (\ref{Fig3}) (a) shows the $I-V$ curves of a pyridine-polyyne-pyridine junction, where the polyyne has $n=3$ carbon pairs.
Following our scheme, we apply a rigid shift to the orbitals of the extended molecule up to and including the nitrogen atoms
at the two pyridine rings. The resulting $I-V$ curve, plotted in Fig. (\ref{Fig3}) (a), shows that the scheme fails calamitously to describe the NDR response of
this junction, even though it improves the equilibrium-based calculation of the $I-V$ curve, which always increases. Furthermore, we have tested that extending the shift up to the inner carbon atoms in the pyridine rings does not improve matters.
In contrast, the scheme provides reasonably accurate curves for similar polyyne-based junctions where the pyridines are replaced by sulfur
atoms (Fig. (\ref{Fig3}) (b)) or by sulfur-benzene rings (Figs. \ref{Fig3}) (c-d)). The corresponding $T(E)$ curves for the pyridine-polyyne-pyridine and
sulfur-benzene-polyyne-benzene-sulfur junctions, calculated with NEGF and with our scheme, are shown in Fig. (\ref{Fig4}). This figure shows how in the first case, a
sharp resonance exists, which moves to higher energies as the bias increases while maintaining its shape. In contrast, our scheme predicts that the resonance
does not shift with voltage because it is placed at the center of the molecule, as we show in Fig. (\ref{Fig4}) (a2). The impact of a finite bias in the second
junction, however, is manifested in the destruction of the HOMO resonance, shown in Fig. (\ref{Fig4}) (b), which is now well captured by our scheme.

The criterion to discern when the proposed scheme can  be used is related to the nature and features of the molecular level responsible for the transport properties.
Those are its spatial placement in the molecule, its energy positioning as well as its coupling to the electrodes. For the pyridine-polyyne-pyridine molecule,
the LUMO orbital is made essentially of carbon $\pi$ states and is located on the central part of the molecule. When the bias is applied, the voltage drop does not
occur at this orbital because the highly conjugated character of the molecule screens the potential. Therefore the applied bias has a small impact in the LUMO orbital
and as a result the shape of the associated transmission resonance does not change. The resonance shifts up in energy because there is a net transfer of charge from
the electrodes towards the molecule. This moves all levels upwards and pins the LUMO orbital to the negative electrode. For the sulfur-benzene-polyyne-benzene-sulfur
junction, the relevant orbital is the HOMO, which is mostly located at the sulfur atoms. The atomic sulfur states couple across the molecule and produce bonding
and anti-bonding states. When the bias is ramped up, those sulfur states follow the chemical potential of each electrode because they are strongly coupled to the leads.
As a consequence, the bonding and anti-bonding molecular levels are destroyed. This is seen in the transmission as a reduction of the HOMO-related resonance, which produces a NDR response. We note that this NDR behavior does not occur if the molecule does not contain the benzene rings as shown in Fig. (\ref{Fig3}) (b).
In this case the coupling between the sulfur states and the carbon chain is much stronger. This increases the hybridization between them, keeps the integrity
of the bonding and anti-bonding states and decreases the impact of the shifts. An intermediate case between these two junctions corresponds to the asymmetric
sulfur-benzene-polyyne-sulfur molecule shown in Fig. (\ref{Fig3}) (c). This junction gives different $I-V$ characteristics for each bias polarity, producing a rectifying response.
In this case the relevant orbital is again the HOMO state, which is mostly localized in the right sulfur atom and the carbon chain. The chain states are left unaffected
when we ramp up the bias and apply the scheme to the sulfur on the right side and the benzene carbon atom closer to the chain. The scheme reproduces quite well both the NDR and the rectification effects.

The scheme proposed is specifically suited for junctions where the states are well localized at the linking atoms or near the electrodes, because then the frontier orbitals are strongly
affected by the applied bias. Large junctions are difficult to simulate using NEGF techniques because they require huge amounts of computer resources and have
frequently convergence problems. Our scheme simplifies the calculation of $I-V$ curves drastically, because it relies on equilibrium techniques, which
are much simpler and computationally much less demanding. We have indeed verified that the $I-V$ curves
corresponding to a series of graphene junctions showing NDR response\cite{Carrascal12} are accurately described with our scheme. The recipe proposed in this article
can be used whenever the ground state at the molecule is not changed by the non-equilibrium conditions. In the electrostatic spin effect, for example, the spin multiplicity of the molecule is changed by the application of a finite bias\cite{Baa09}, and as a result our scheme cannot be applied.

In summary, we propose in this article that current-voltage characteristics of a wide array of electronic junctions can be accurately described by a scheme that
combines conventional equilibrium transport techniques with suitable shifts of energy levels up to the weak links existing in those junctions. This scheme can
have drastic practical effects because equilibrium codes are much simpler and computationally more efficient than non-equilibrium codes. We have corroborated very large
computational speed-ups in the above examples. As a consequence, our recipe opens a route towards realistic transport simulations
comprising much bigger or complex systems than those that could be feasibly simulated with current NEGF techniques.

The research presented here was funded by the Spanish MICINN through the grant FIS2009-07081 and by the Marie Curie network nanoCTM.
VMGS thanks the Spanish Ministerio de Econom\'{\i}a y Competitividad for a Ram\'on y Cajal fellowship (RYC-2010-06053).


\begin{thebibliography}{References}
\bibitem{NDR96}
NDR from equilibrium transport properties has already been observed in model molecular wires using a mean-field Hubbard model\cite{Muj96}. The NDR was produced by voltage-induced localization.
\end{thebibliography}
\end{document}